# Re-examining rates of lithium-ion battery technology improvement and cost decline

Micah S. Ziegler[†]     and     Jessika E. Trancik[†,‡,*]

August 2020

## Key numerical results

- Previous studies have reported wide-ranging annual price decrease percentages for lithium-ion technologies between 8.8 and 29% and learning rates between 14 and 30% leading to an ambiguous assessment of improvement rates for a critical low-carbon technology.

- Simple projections based these previous assessments' price data would suggest lithium-ion cells are on a path to reach a 75 USD/kWh threshold anytime between 2009 and 2027 and a 20 USD/kWh threshold anytime between 2015 and 2042. While any forecasting model should not rely solely on a simple extrapolation of past trends, this comparison demonstrates the significance of the differences in reported data.

- We revisit and reconcile the diverse data sets that underlie these analyses and transparently define series to represent each type of data. Based on these series, which are more comprehensive than the datasets previously employed, we estimate that the price of lithium-ion cells has declined by about 97% since their commercial introduction in 1991.

- We estimate that between 1992 and 2016 energy capacity–scaled real prices declined by an average of 13% per year for both all types of cells and cylindrical cells, while learning rates are estimated to be 20% for all cell types and 24% for cylindrical cells. In addition, a 40% decline in price was observed with each doubling of cumulative inventive activity as measured by patent filings.

- Expanding the definition of service to include energy density results in a faster measured rate of technological improvement. Measured annual declines in price per service increased from 13 to 17% for both all technology types and cylindrical cells. Learning rates increased similarly, from 20 to 27% for all cell shapes and 24 to 31% for cylindrical cells.

[†]Institute for Data, Systems, and Society, Massachusetts Institute of Technology, Cambridge, MA, USA.
[‡]Santa Fe Institute, Santa Fe, NM, USA.
[*]trancik@mit.edu

# Abstract

Lithium-ion technologies are increasingly employed to electrify transportation and provide stationary energy storage for electrical grids, and as such their development has garnered much attention. However, their deployment is still relatively limited, and their broader adoption will depend on their potential for cost reduction and performance improvement. Understanding this potential can inform critical climate change mitigation strategies, including public policies and technology development efforts. However, many existing models of past cost decline, which often serve as starting points for forecasting models, rely on limited data series and measures of technological progress. Here we systematically collect, harmonize, and combine various data series of price, market size, research and development, and performance of lithium-ion technologies. We then develop representative series for these measures, while separating cylindrical cells from all types of cells. For both, we find that the energy capacity–scaled real price of lithium-ion cells has declined by about 97% since their commercial introduction in 1991. Using performance curve models, we estimate that between 1992 and 2016, real price per energy capacity declined 13% per year for both all types of cells and cylindrical cells, and upon doubling cumulative market size, decreased 20% for all types of cells and 24% for cylindrical cells. We also develop a method to incorporate additional performance characteristics into these models, including energy density and specific energy performance metrics. When energy density is incorporated into the definition of service provided by a lithium-ion battery, estimated technological improvement rates increase considerably. The annual decline in real price per service increases from 13 to 17% for both all types of cells and cylindrical cells while learning rates increased from 20 to 27% for all cell shapes and 24 to 31% for cylindrical cells. These increases suggest that previously reported improvement rates might underestimate the rate of lithium-ion technologies' change. Moreover, our improvement rate estimates suggest the degree to which lithium-ion technologies' price decline might have been limited by performance requirements other than cost per energy capacity. These rates also suggest that battery technologies developed for stationary applications, where restrictions on volume and mass are relaxed, might achieve faster cost declines, though engineering-based mechanistic cost modeling is required to further characterize this potential. The methods employed to collect these data and estimate improvement rates are designed to serve as a blueprint for how to work with sparse data when making consequential measurements and forecasts of technological change.

# Broader context

Energy storage technologies have the potential to enable greenhouse gas emissions reductions via electrification of transportation systems and integration of intermittent renewable energy resources into the electricity grid. Lithium-ion technologies offer one possible option, but their costs remain high relative to cost-competitiveness targets, which could hinder these technologies' broader adoption. Existing measures of the rate at which lithium-ion technologies' costs have fallen differ considerably, resulting in an ambiguous assessment of their past improvement rates. We collect and harmonize data that describe how lithium-ion technologies have improved and possible drivers of their advancement. We use performance curve models to measure how lithium-ion technologies have changed over time as well as with increasing market size and inventive activity. In addition, we present a method to incorporate other dimensions of performance into our measures of technological change, allowing us to also consider increases in energy density and specific energy. Our results begin to approximate how previous measures might have underestimated the rate of lithium-ion technologies' improvement and suggest how much faster these technologies might advance when other characteristics are prioritized. Moreover, we delineate methods that can be applied to study how these and other energy and environmentally relevant technologies change over time, to refine efforts to inform public policies, investments, and technology development.

## Introduction

Energy storage can help enable renewable energy adoption and greenhouse gas emissions reductions. Toward these goals, electrochemical energy storage technologies are increasingly employed to both electrify transportation systems and aid electricity production and grid reliability.[1–3] While these storage technologies have the potential for substantially wider adoption, their costs remain relatively high, especially in comparison to cost-competitiveness targets absent a robust price on greenhouse gas emissions.[4–9] As such, considerable interest exists in modeling how storage technologies' costs change over time and which research directions, business strategies, and policy incentives could help lower these costs.[4,6,8,10–14] Increasingly, many researchers, technology developers, and electricity providers have focused on lithium-ion technologies, whose historic cost decline has been cited as a significant achievement and promising trend.[2,15–18] However, uncertainty remains as to the rate at which lithium-ion technologies' costs and prices have fallen, adding to uncertainty about the potential for their continued decline.[10,11,14,19–21] In addition, there is growing recognition that characteristics beyond energy capacity cost, including cycle-life and capacity-loss characteristics, could influence the adoption of energy storage technologies.[4,20,22]

The need for better characterization of technological improvement rates applies to many technologies. Technologies are constantly changing and especially for those expected to help enable climate change mitigation, such as energy storage, it is important for society to be able to accurately measure and interpret estimates of their rates of technological change. In this paper, we return to this challenge. We carefully examine the case of lithium-ion battery technologies, with the goal of better characterizing improvement rates for these technologies and developing a more general blueprint that can be applied to other technologies.

To analyze the rates of energy storage systems' cost declines, some researchers and industry analysts have turned to phenomenological models of cost change.[23–30] These models are often exponential or power relationships between the cost or price of a technology and possible determinants, such as: time, production quantity, proxies for research and development activity, or a combination of these variables.[27] The rates estimated from these analyses are then sometimes used to project future cost changes, especially how a technology's cost could decline as its production is increased, though such projections should always be accompanied by an error model.[27] Over the past few decades, this approach has been employed to study and forecast cost reduction for variety of climate-relevant energy technologies,[31–38] such as photovoltaic panels and wind turbines. More recently, similar analyses have been performed for energy storage technologies, with a focus on lithium-ion batteries for both mobile and stationary applications.[12,14,21,39–49] These analyses have primarily examined the relationship between the historical price of lithium-ion cells (typically in terms of price per energy capacity, such as USD per kWh) and cumulative production (in total energy capacity, in units of MWh) and derived rates of price decline, often denoted "learning rates."[a] These learning rates represent the price decline observed upon a doubling of cumulative production capacity and are often employed to project further price declines from increased production. Additional work has detailed the uncertainty associated with projections based on these rates and highlighted the need for error models to accompany forecasts based on past trends.[27,50]

Analyses of price versus production for small lithium-ion cells have estimated a wide range of learning rates, spanning 14 to 30%.[46,51] Simple projections based on this range of rates arrive at widely varying conclusions as to when lithium-ion technologies might cross cost or price targets and the associated investment required. Such sensitivity of target dates and investment requirements to small changes in technology improvement rates is described by the nonlinear power law and exponential relationships. Variations in the datasets for other energy technologies have led to similar discrepancies between retrospective analyses, highlighting the importance of reducing data uncertainty.[26,38,50,52–54] Moreover, nearly all of these analyses focus on one performance metric of lithium-ion technologies: the cost or price per energy capacity. However, since their commercial introduction lithium-ion technologies have improved along many dimensions of performance, notably packing more energy and power into cells, expanding their utility in a variety of applications.[17,55–58] Despite being prominent objectives of research and development and drivers of technological adoption, these physical performance improvements are often considered separately from cost and price declines, possibly distorting estimates of technological improvement rates.

A clear understanding of past trajectories can help to determine reliable measures, rates, and directions of technological improvement, as well as estimates of uncertainty in data and forecasts, for lithium-ion and other technologies. In addition, reliable estimates of historical trends are a key component of mechanistic models of cost change, which seek to elucidate the impact individual factors have on a technology's overall cost and explain how a technology's cost has changed in an effort to inform future reductions.[59,60] In this work, to develop improved estimates of the rates of lithium-ion technologies' change, we collect, harmonize, and combine multiple historical data sources describing the price, production, and development of lithium-ion technologies. We then systematically develop representative data series that estimate how lithium-ion technologies and their proposed drivers have changed over time, in the process transparently outlining the definition of "representative" so that it might be adapted or improved as required to answer other research questions. When possible, data are split into subgroups based on cell shape, with a focus on separating cylindrical cells from all cell shapes. We then explore the relationships between the price decline of lithium-ion technologies and a range of factors, including time, market size, and research and development activity. Throughout, we delineate our analyses to enable fair comparison between various models of technological improvement and with previously published results.

We also consider other characteristics of lithium-ion cells that have changed over time, notably energy density and specific energy, both of which have improved substantially. We propose a method to expand the definition of service provided by a lithium-ion cell to include multiple characteristics. We then develop performance curves that represent how these physical characteristics have changed over time and use these curves along with the representative price series to explore how lithium-ion technologies have improved more broadly. We find that incorporating these additional characteristics considerably increases estimated rates of technological change, suggesting that these technologies have improved faster than estimated based on cost metrics that do not account for the service improvement that higher energy density and specific energy have provided for some applications, such as in mobile devices and electric vehicles. Overall these results provide a more complete picture of the actual rate of past improvement of lithium-ion technologies and begin to suggest that faster cost improvement may be possible in the future for applications with relaxed volume and mass restrictions, as in the case of stationary energy storage.

# Methods

## Data series collection

We collected data from articles, reports, and presentations from the academic, governmental, and business literature with a focus on tracing data as far as possible to the original source. Original data were sought in order to improve data and metadata quality and reduce the chance of double-counting data points. For example, a variety of recent performance metric series were excluded as their underlying data could be obtained and used directly.[14,15,32,42,46,51,61–67] However, data series that combined previously reported data with otherwise unreported data were included. Similarly, we included data series reported with unclear references or assumptions, even if the data series closely resembled a series reported earlier. When a researcher or organization presented the same series or updated versions of a given series over the course of multiple presentations or reports, the most recent available data were incorporated. Modeled estimates of cost were excluded from this analysis. Patent filing counts data series were acquired from multiple patent databases.[68,69]

When developing representative series employed in this analysis, data series that were clearly derived from other sources included in our analysis were excluded to prevent overreliance on those data. Additional details on the collection and harmonization of the data series employed in this analysis can be found in the supplementary information (SI).

## Lithium-ion technology database

Data on individual battery specifications and prices were collected from a variety of academic, government, industrial, and commercial sources and compiled into a human- and machine-readable database. The database contains 1716 unique records of cells employing lithium-ion and lithium-ion polymer technologies for the years 1990 through 2019. Additional details describing the development and structure of the database, as well as how energy density and specific energy values were calculated from other reported metrics, are available in the SI.

## General computational methods

Currency conversion, inflation adjustment, database parsing, and plotting were performed using R (v 3.6.2).[70] String manipulation and comparison were implemented using stringi.[71] Data series and the database of battery performance metrics were stored in Microsoft Excel files (xlsx format) and read and modified in R with the help of the readxl[72] and openxlsx[73] packages. Conversion of calendar dates to decimal dates for use in modeling was performed using lubridate.[74]

## Modeling

Models of relationships between the prices of lithium-ion cells and various determinants were calculated by first taking the base-10 logarithm of the price series and determinant values, if appropriate, and then performing a linear regression. Linear regressions were performed using the ordinary least squares method via R's *lm* function, and the resulting $R^2$ values reported herein are adjusted. Unless otherwise stated, shaded regions plotted alongside trend lines are prediction intervals calculated at the 0.95 level using the *predict* function in R.

## Currency conversions

Historical foreign exchange rates for the conversion of Japanese Yen and Australian Dollars (AUS) to US Dollars (USD) were obtained from the Board of Governors of the Federal Reserve System.[75] The Yen to USD dataset includes yearly, monthly, and daily rates, all released on 2020-06-01. The AUS to USD dataset comprises yearly rates, released on 2020-06-15.

## Inflation adjustment

Unless otherwise noted, nominal currency values in US Dollars were adjusted for inflation using the Implicit Price Deflator for Gross Domestic Product (table 1.1.9) published by the US Department of Commerce's Bureau of Economic Analysis,[76] as has been employed previously.[77] The dataset was revised on 2020-05-28 and contains series with both yearly and quarterly resolution. Quarterly resolution GDP deflator values were employed to adjust monthly data series.

## Limitations

While we strove to collect data from a wide variety of physical and digital sources, searching for and reading of references was primarily conducted in English. When potentially useful resources were encountered in other languages, translation relied on various online tools (*e.g.* Google Translate).

## Results

### Development of representative price, market size, and patent filing data series

Researchers and analysts have reported a variety of analyses that regress the cost or price of lithium-ion technologies against proposed cost change determinants, resulting in a wide range of proposed improvement rates for lithium-ion technologies (Table S2). Their models examine the decline in cost or price at the cell, pack, and system levels and explore the decline's relationship to determinants including production, inventive activity, time, and material prices. To demonstrate the diversity of data underlying these analyses of lithium-ion technologies' improvement, we collected and harmonized as many distinct data series as possible. In these efforts, we strove to obtain data directly from their original sources and systematically investigated whether data were adjusted for inflation or converted from one currency to another (additional details available in Methods and SI). To reconcile the differences between data series, we categorize them, and within each category, we transparently define and construct a "representative" series from the individual data series. These representative series are designed to incorporate the most reliable data available and cover as many years of technology development as possible. We detail the approaches taken to develop these representative series to clarify and mitigate the impact of data uncertainty. We expect these approaches and the resulting series can be improved over time as new data become available and to answer different research questions.

Our data collection yielded 25 series that track lithium-ion cell cost or price change over time. Series were converted to energy capacity–scaled real costs or prices in units of 2018 USD/kWh and are presented along with similarly harmonized single-year records of cell-level prices (Figure 1). Some single-year records are of cell purchases by academic researchers, which are typically much higher than industry-wide price estimates, likely due to price markups associated with ordering small numbers of cells. While cost data are typically preferred for phenomenological studies of cost change,[33,50] empirical price data were much more commonly reported for lithium-ion technologies than cost data were, as has been observed previously.[42,51,78] Taken together, the data reveal a consistent decrease in lithium-ion cell price over time, with a few exceptions around 1995 and 2008. Overall, prices have declined by about 97% since the commercial introduction of lithium-ion cells in 1991.

Fitting these price series with negative exponential growth (decay) curves with time results in a wide range of estimated annual price decrease percentages, from 4.8 to 23% (Figure 1). A similarly wide range of percentages (8.8–29%) is observed when examining the price series specifically employed in previous analyses of the relationship between the price of lithium-ion cells and cumulative production, installation, or sales, as measured in units of energy capacity (Figure 2, Table S2). In turn, this wide variation in rates of price decline considerably impacts the estimated learning rates (14–30%) and projections of when prices are expected to meet certain targets, yielding crossover ranges that span decades (Figure 2 and Figures S2–S5).

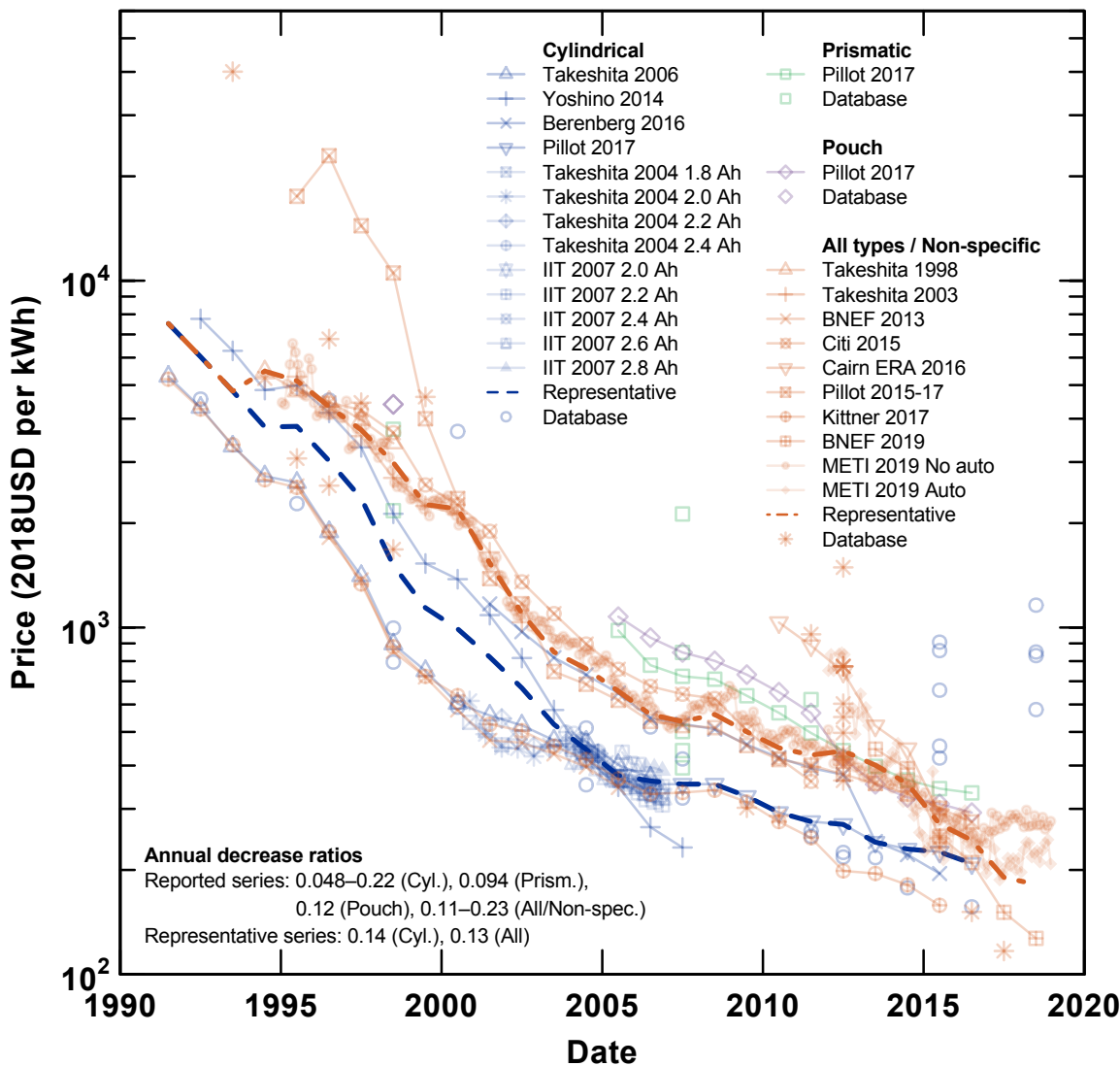

Figure 1: Lithium-ion cell prices. Time series and single-year records of lithium-ion cell prices for cylindrical (blue), prismatic (green), pouch (purple), and all types (orange) of cells, as well as representative price series for cylindrical (blue, bold, dashed) and all types (orange, bold, dashed) of cells. Records that did not specify cell type are included with series representing all types of cells. Series specifically describing cylindrical cells have annual decrease ratios between 0.048 and 0.22 while those describing all types of cells have ratios that span 0.11 to 0.23. The representative series of cylindrical cell prices has an annual decrease ratio of 0.14 (for 1991 through 2016) while that for all types of cells has a ratio of 0.13 (for 1991 through 2018). A version of this plot with a non-logarithmic dependent axis is included as Figure S1.

Many factors contribute to the diversity of price data and rates of price decline. Notably, substantial price differences can be observed between different system levels and cell shapes. As such, we sought to only combine data series that describe technologies with the same design. For example, one can distinguish between lithium-ion cells, modules, packs, and systems.[14] In this work, we focus on cells and attempt to further differentiate the group of lithium-ion cells based on cell shape. Lithium-ion cells are manufactured in a variety of shapes, the three most prominent being cylindrical, prismatic, and pouch. The earliest cells were cylindrical,[58,79–81] and prismatic and pouch cells were introduced later.[82–84] The price data indicate that cylindrical cells are on average less expensive for a given energy capacity than prismatic or pouch cells are. Examination of the harmonized data (Figure 1) reveals two major groups of price data series. The first group contains series specific to cylindrical cells and series reported without specifying a cell shape (*i.e.* "non-specific" series) but that appear to be derived from cylindrical-specific series. These derived series were identified by their very close similarity to cylindrical-specific series that had been published previously. The second group contains price series averaging across all cell shapes, including many other non-specific series. Careful parsing of the data allowed us to develop two representative price series, one representing the price for cylindrical lithium-ion cells and another for all cell types (Figures 1–2). Generally, these representative price series were developed by combining series comprising industry-wide cell-level price per energy capacity estimates and averaging concurrent portions of these series. Single-year price estimates were employed to corroborate these series' data but not incorporated into the averages. Additional details are provided in the SI.

In addition to grouping data by the represented technology's design, we also found that price data can be distinguished by a cell's intended application. Notably, lithium-ion technologies can be differentiated based on whether cells were designed and manufactured for use in portable electronics versus those destined for automotive applications.[44,85] While most data series did not provide such a distinction, it is observable in the data obtained from Japan's Ministry of Economy, Trade and Industry (METI), which starting in 2012 separated batteries for use in automobiles from those for use in other applications (Figure 1).[86] Between 2012 and 2015, the decline in price of these automotive cells was considerably greater than the price decline observed in most other series and mirrors the sharp declines in the series reported by Cairn Energy Research Advisors (Cairn ERA) in 2016[87] and Bloomberg New Energy Finance (BNEF) in 2019.[11] Since 2015, the METI data series suggest that the price change in automotive cells has been more gradual. A series including METI data on both types of lithium-ion cells was employed in the development of the aforementioned representative series for all cell types.

We similarly collected and harmonized 27 data series recording the size of the market for lithium-ion cells over time, measured in number of cells per year (Figure 3). Data series included production, shipment, sales, and demand data, and were relatively consistent with each other. However, the sum of data collected from Japanese, Korean, and Chinese government resources (see Figure S6, and SI for references) provide a higher estimate of cells produced than other sources suggest, especially between 2015 and 2017. Combining reliable data sources yielded representative series for cylindrical cells, cylindrical and prismatic cells combined, and all cell types (Figure 3). The representative series for the market size of cylindrical cells was constructed similarly to the price series, with multiple series being combined and averages taken where series overlapped. Considering the divergence observed in the all cell types series and the reliability of different sources, when concurrent data series disagreed on the market size of all cell types, the maximum value was employed. (Additional details are provided in the SI.) These representative series indicate that since 1992, the market for cylindrical cells has grown by about 3.5 orders of magnitude, while that for all types of cells has grown 4.1 orders of magnitude.

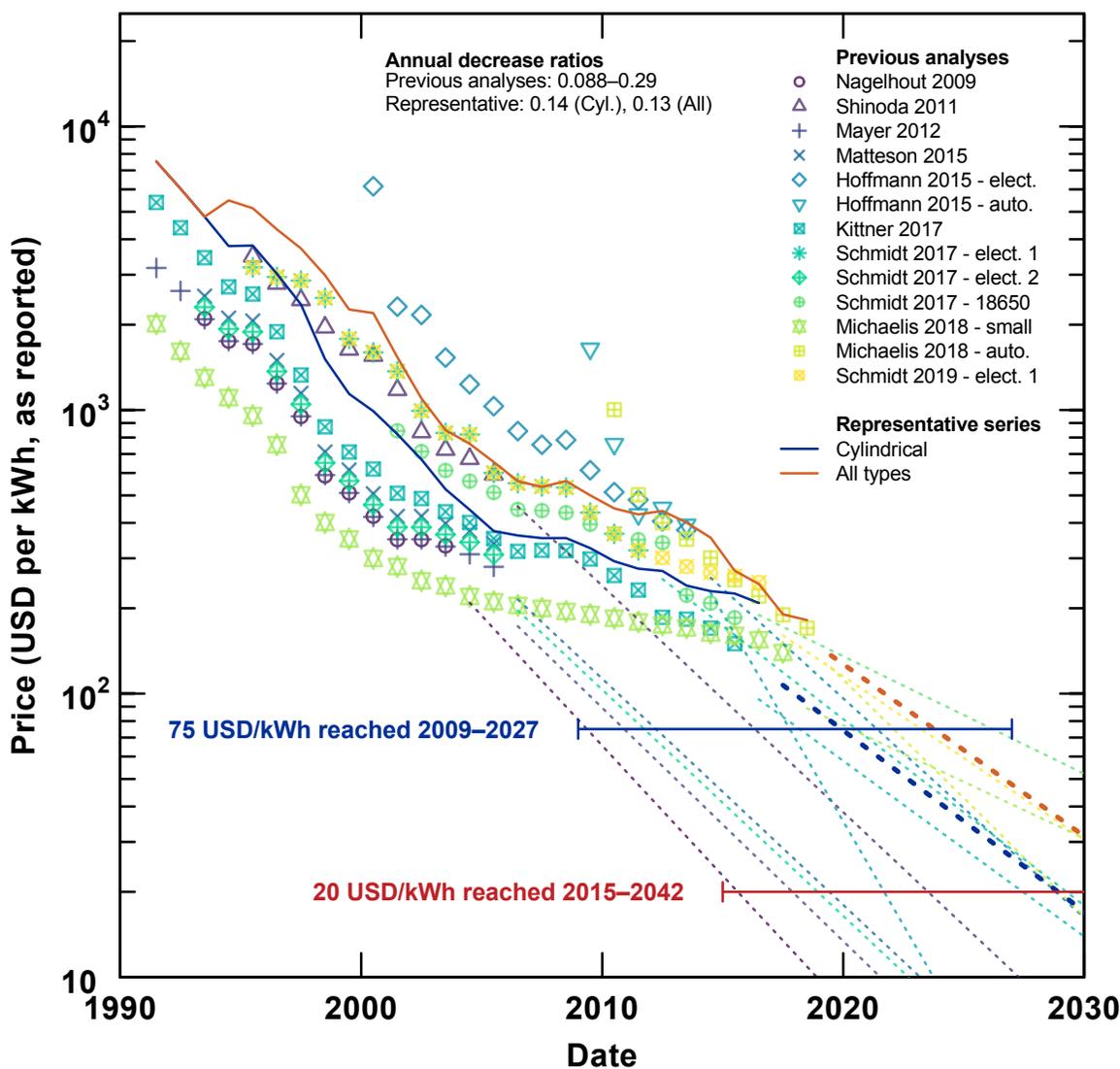

Figure 2: Reported lithium-ion cell price series and projections based on simple extrapolation to demonstrate the consequences of data uncertainty. The lithium-ion cell price per energy capacity series included here were used in previously reported performance curve analyses of cell-level price vs cumulative market size (*e.g.* production, installation, sales, etc.) as measured in energy capacity. The representative series are developed in this work (*vide supra*). Modeling the price versus time data series as exponential declines provides estimates of annual decrease ratios which in turn are used to develop the projections. These simple projections, which are intended to examine the differences in the underlying data, suggest a nearly 20-year range for when prices might cross a 75 USD/kWh threshold and a nearly 30-year range for reaching 20 USD/kWh. Additional methodological details are available in the SI, along with price projections based on market size projections (Figures S2–S5).

Market size data in units of energy capacity (MWh) were similarly collected, and a representative series for all cell types was developed (Figure 4) by combining multiple series and averaging reliable concurrent data. Generally both measures of market size are consistent; they indicate a rapid growth in annual market size between 1991 and 1996, followed by slower growth from 1997 onward. However, the recent uptick in market growth observed for all cell types as measured in number of cells is not reflected in market size estimates measured in energy capacity. The representative series developed in units of MWh suggests an increase in market size of nearly six orders of magnitude since 1991 and about 4.7 orders of magnitude since 1992.

In addition, data on the annual filings of simple patent families associated with lithium-ion technologies were collected from Google Patents[68] and Patsnap's[69] databases using an International Patent Classification symbol specific to lithium-ion batteries. The resulting series (Figure 5) are generally consistent with those reported by Mayer et al.[51] and Kittner et al.[12] Nearly all series also display a sharp drop in patent counts in their last year, very likely reflecting mid-year data collection or delays between patent filing and publishing and database updating.[88] As many of the advancements in lithium-ion technologies can be incorporated into cells regardless of their shape, patent filings were not divided into shape-specific subgroups. In this work, the series of data obtained from the PatSnap database is used as the representative series for annual patent filings. This series indicates an increase of simple patent family filings of nearly four orders of magnitude since 1977. The observed increase in patent filings serves as a proxy in this analysis for the growth in research and development activity directed at improving lithium-ion technologies, which is consistent with increased production of cells and the cells' use in an expanding range of energy storage applications.

## Analyses of relationships between price and determinants

A variety of performance curve models can be employed to examine the degree to which the price of lithium-ion technologies is correlated with possible determinants, and thus gain insight on the relationships between variables in phenomenological models. These models can provide a top-down view of cost reduction trends and determinants and complement bottom-up, mechanistic modeling approaches that seek to disentangle cost contributors in an effort to explain cost decline.[59] The models tested include:

$$\log(y_t) = at + b + e_t \tag{1}$$

$$\log(y_t) = a\log(x_t) + b + e_t \tag{2}$$

$$\log(y_t) = a\log(q_t) + b + e_t \tag{3}$$

$$\log(y_t) = a\log(z_t) + b + e_t \tag{4}$$

$$\log(y_t) = a\log(v_t) + b + e_t \tag{5}$$

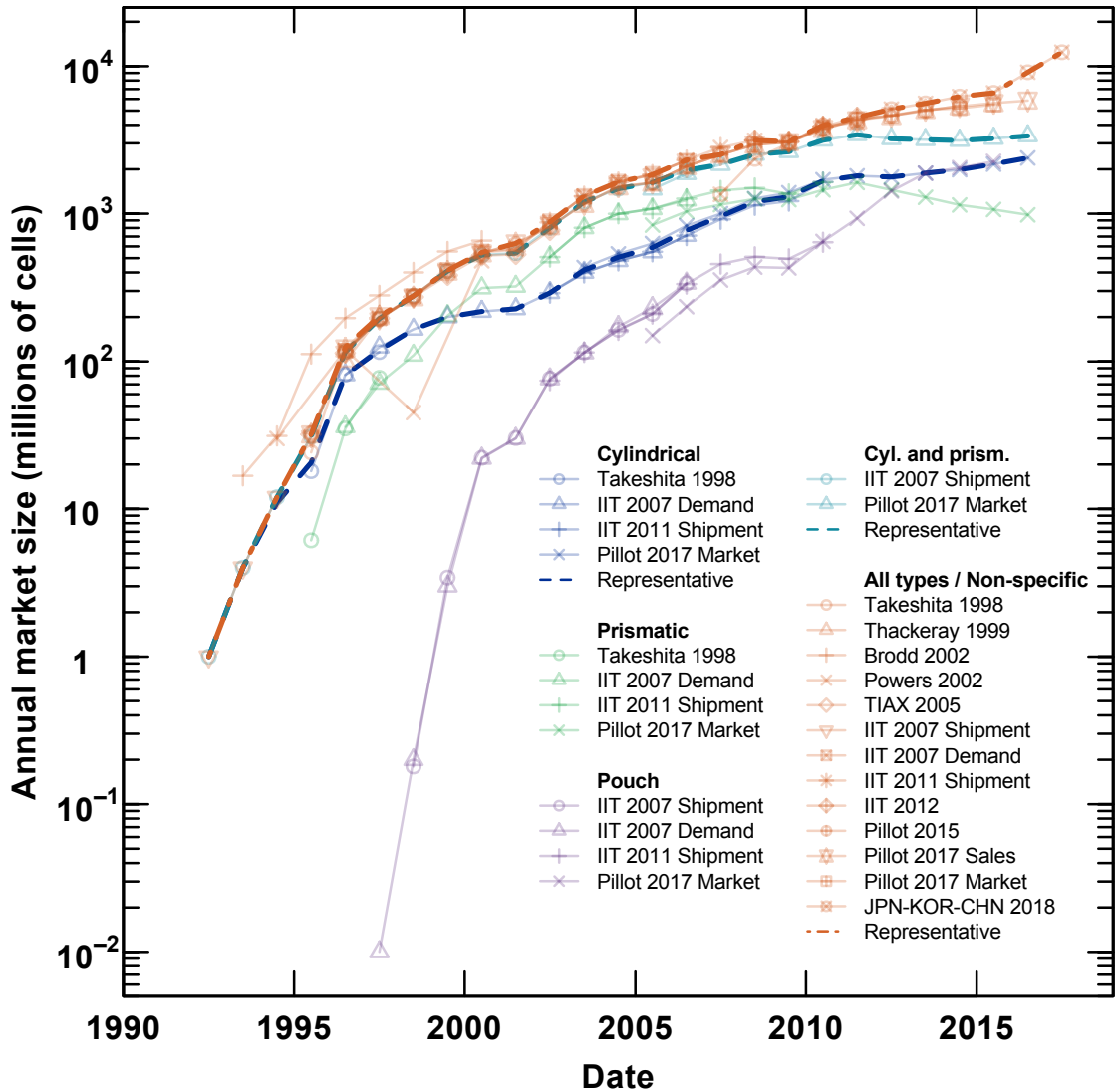

Figure 3: Lithium-ion market size measured by number of cells. Time series of lithium-ion market size measured in number of cells for cylindrical (blue), prismatic (green), pouch (purple), cylindrical and prismatic (light blue), and all types (orange) of cells, as well as representative price series for cylindrical (dark blue, bold, dashed), cylindrical and prismatic (light blue, bold, dashed) and all types (orange, bold, dashed) of cells. Records that did not specify cell type are included with series representing all types of cells.

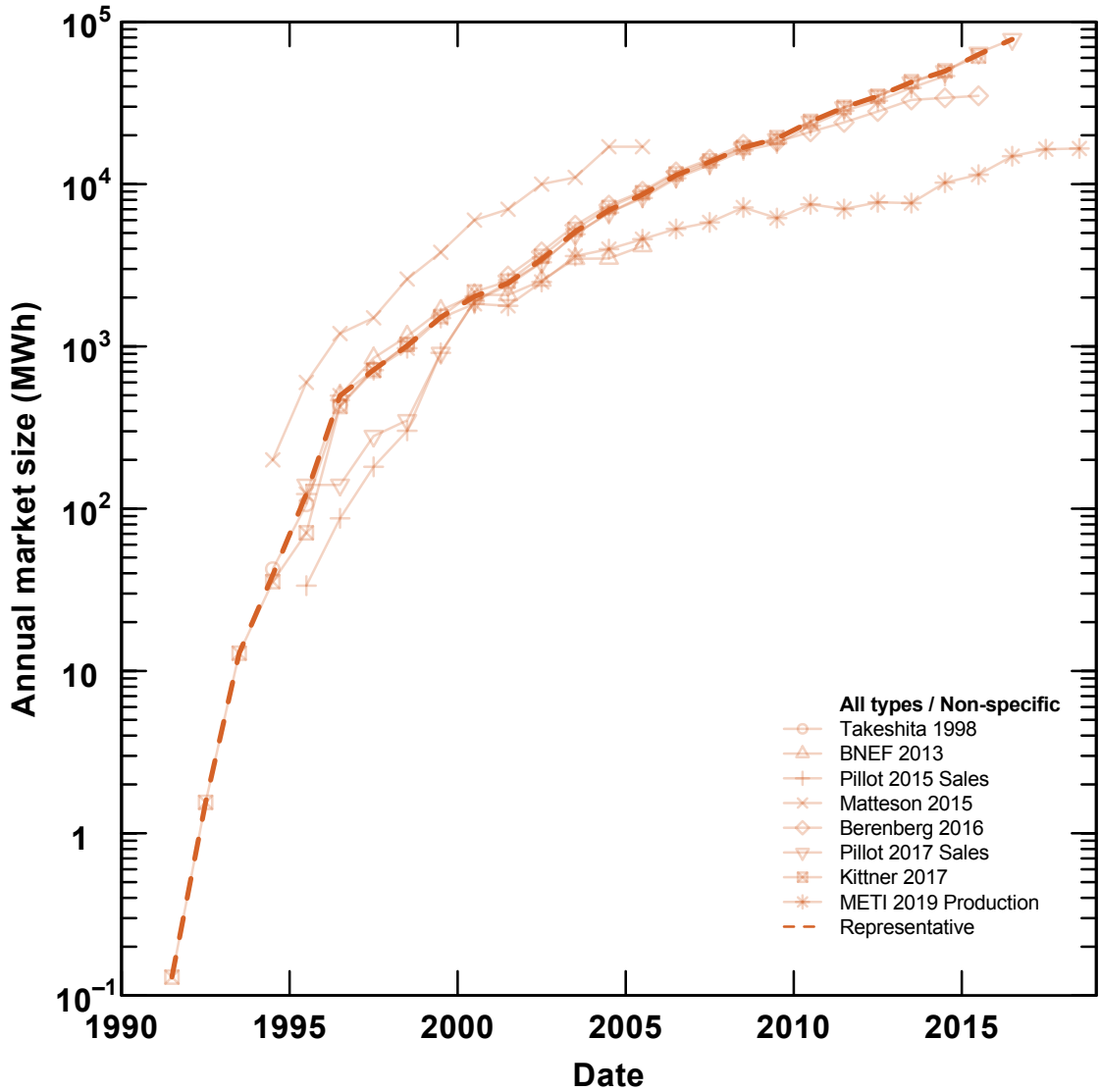

Figure 4: Lithium-ion market size measured by cell energy capacity. Time series of lithium-ion market size measured in aggregate cell energy capacity for all types of cells, as well as a representative market size series (orange, bold, dashed) of cells. Records that did not specify cell type are included with series representing all types of cells.

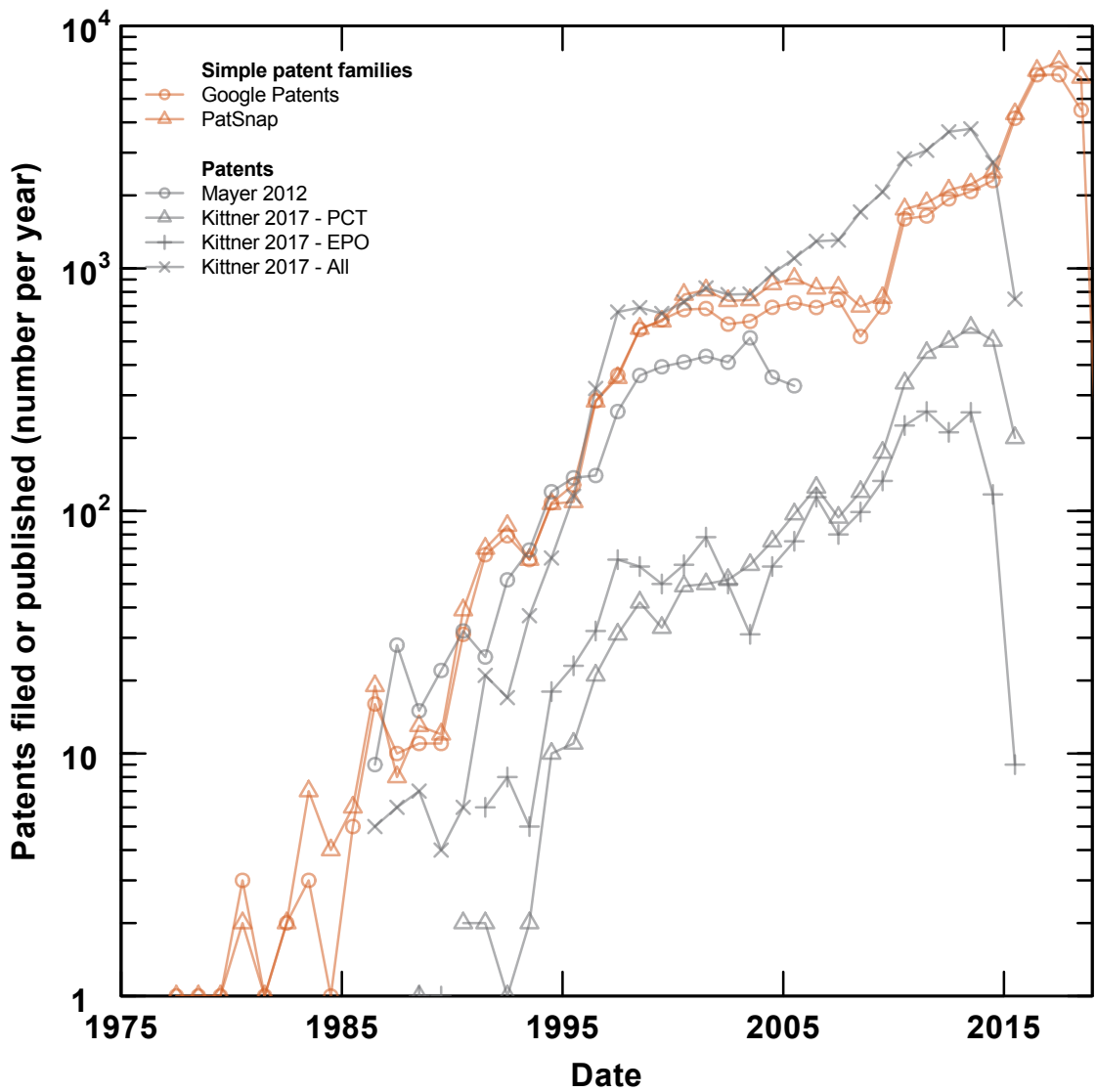

Figure 5: Annual growth in lithium-ion–related patent filings. Time series of the number of lithium-ion technology patent filings between 1975 and 2018 both developed in this work (orange) and previously reported (gray).

In all models, $y_t$ is a measure of technological progress or performance, which in the case of energy storage technologies is typically represented by the real price of cells, packs, or systems scaled by their energy capacity (*i.e.* 2018 USD/kWh). The first model (eq. 1) suggests that a technology improves (*e.g.* scaled real cost or price declines) with time and is colloquially known as Moore's law.[89] The second two models (eqs. 2 and 3) examine the relationships between scaled real cost or price and cumulative production ($x_t$) and annual production ($q_t$), commonly referred to as Wright's[29] and Goddard's[90] laws, respectively. In this work, market size is used as a proxy for production, which is consistent with the good agreement between production, sales, demand, and market size data series (*vide supra*). The final two equations (eqs. 4 and 5) examine how a technology changes with cumulative ($z_t$) and annual ($v_t$) research and development activity. In this work, annual patent filing counts are employed as a proxy for research and development (*i.e.* "inventive" or "innovation") activity, an approach that has been supported by studies of other energy technologies.[88,91,92] Constants ($a$ and $b$) and the residuals ($e_t$) differ for each model.

These models are used to measure improvement rates that are commonly employed when comparing results of performance curve analyses or projecting future improvements. In the case of equation 1 and employing a base-10 logarithm, the annual decrease ratio (ADR) is given by

$$\text{ADR} = 1 - 10^a \tag{6}$$

In the case of equation 2, the learning rate (LR) is defined as:

$$\text{LR} = 1 - 2^a \tag{7}$$

and is comparable to many of the "experience" and "learning" rates previously reported. The learning rate represents the decrease in cost or price projected for a doubling of cumulative market size. In the case of equation 4, an analogous rate, herein referred to as an inventive activity rate (IAR), can be calculated to provide the decrease in cost or price associated with a doubling of research and development activity,

$$\text{IAR} = 1 - 2^a \tag{8}$$

We refer to these three rates (*i.e.* ADR, LR, and IAR) generally as "improvement rates".

When performing these analyses, we employed price and determinant data that describe the same group or subgroup. For example, the price of all types of cells is regressed against the market size of all types of cells, while the price of cylindrical cells is regressed against the market size of cylindrical cells. Patent data could not be easily separated into inventions that only applied to cylindrical cells as opposed to all types of lithium-ion cells because many inventions could apply to both cell designs. Thus, inventive activity rates were only estimated for all types of lithium-ion cells. In addition, to fairly compare the models' results, we generally limit these analyses to the time period for which representative series values were available for both all types of cells and cylindrical cells: the years from 1992 through 2016. As additional reliable data become available, this range can be extended.

For the years between 1992 and 2016, the relationship between the all-cell-types representative price and time, cumulative market size, or cumulative patent filings is measured via application of equation 1, 2, or 4, respectively (Figure 6). The results reveal reasonably high correlations for all three models, as indicated by the coefficients of determination ($R^2$). Meanwhile, correlations between price and annual market size, measured in number of cells, and annual patent filings (*i.e.* equations 3 and 5) are a bit lower (Figures S7 and S8). Compared to the results for all types of cells, a slightly lower correlation is found between cylindrical cells' price and time, while a higher correlation is observed between cylindrical cells' price and cumulative market size (Figure 6).

The observed rate of scaled price decline versus time is very similar for both representative series, with an annual decrease ratio of 13.1% for all types of cells and 13.3% for cylindrical cells. However, the learning rates determined from the association between scaled price and cumulative market size differ substantially, with 20.4% for all types of cells and 24.0% for cylindrical cells. Meanwhile, examination of how scaled price varies with cumulative patent filings provides an estimated inventive activity rate of 40.1%.

Many of the previously reported improvement rates for lithium-ion technologies were determined by applying equation 2 (Wright's Law), regressing price per energy capacity against cumulative energy capacity production to measure learning rates (Table S2). To investigate the variability observed in previous learning rate estimates, we used the representative price per energy capacity and energy capacity market size series for all types of cells to estimate the learning rate for every possible interval of seven or more years between 1991 and 2016 (Figure 7). (This analysis includes 1991 to allow for a fair comparison with previously reported analyses because a few also extend back that far.) The results reveal that even with a single price and market size data series, a wide range of learning rates can be estimated depending on which time period is chosen. As the interval examined lengthens or more recent intervals are employed, the dispersion of learning rates narrows. However, there is no clear trend in the average learning rate as more recent data are employed, unlike the negative trends in learning rate versus interval recency that have been observed in some cases.[93] To encompass a broader range of possible analyses, we similarly examined learning rates that could be estimated by applying equation 2 to cylindrical cell prices and the same energy capacity market size series (Figure S9), even though this market size series is not specific to cylindrical cells. The range of possible learning rates calculated in this fashion for both all types of cells and cylindrical cells encompasses nearly all previously reported learning rates (Figure S10).

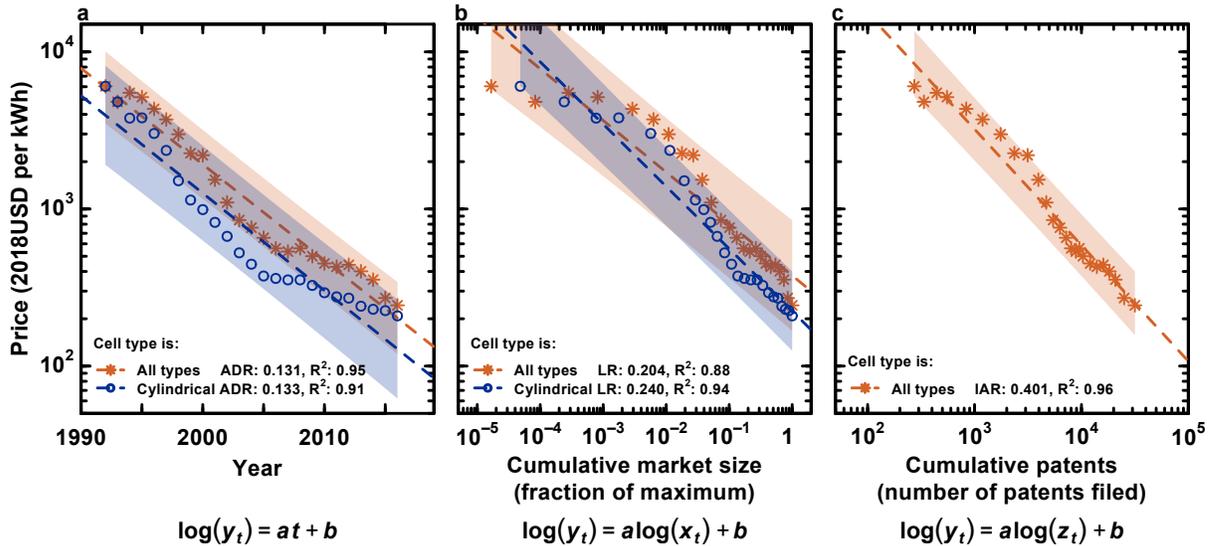

Figure 6: Lithium-ion price per energy capacity regressed against a variety of determinants. Lithium-ion cell price per energy capacity regressed against year (a), cumulative market size (b), and cumulative patent filings (c) for all (orange asterisk marks) and cylindrical (blue circles) cell shapes. Prediction intervals (95% level) are plotted as similarly colored shaded regions. Analyses are restricted to the years 1992 through 2016, for which data are available for both cylindrical and all-cell-types representative price and market size series. Market size for both all types and cylindrical cells is measured in number of cells produced.

We also explored the impacts that different types of market size estimates have on learning rates. Nearly all published learning rates for lithium-ion technologies rely on cumulative market size estimates measured in energy capacity (*e.g.* MWh) as opposed to number of cells. However, annual energy capacity market size values reflect both the number of cells produced and the energy capacity per cell. Energy capacity for a given cell size has increased as lithium-ion technologies have improved but this trend could itself be considered a consequence of research and development, additional production experience, and other activities. In addition, as lithium-ion technologies have expanded into more varied applications, smaller cells have been produced, such as pin- and button-type batteries, leading to cells with lower energy capacity per cell. As such, energy capacity per cell is not necessarily a driver of cost decline in the traditional learning-by-doing or economies-of-scale models, especially when focusing on production processes, and instead could be considered a measure of technical performance. To explore the impact of market size measurement type on learning rates, equation 2 was employed to calculate learning rates and their errors[50] using different measures of cumulative market size (Table 1).

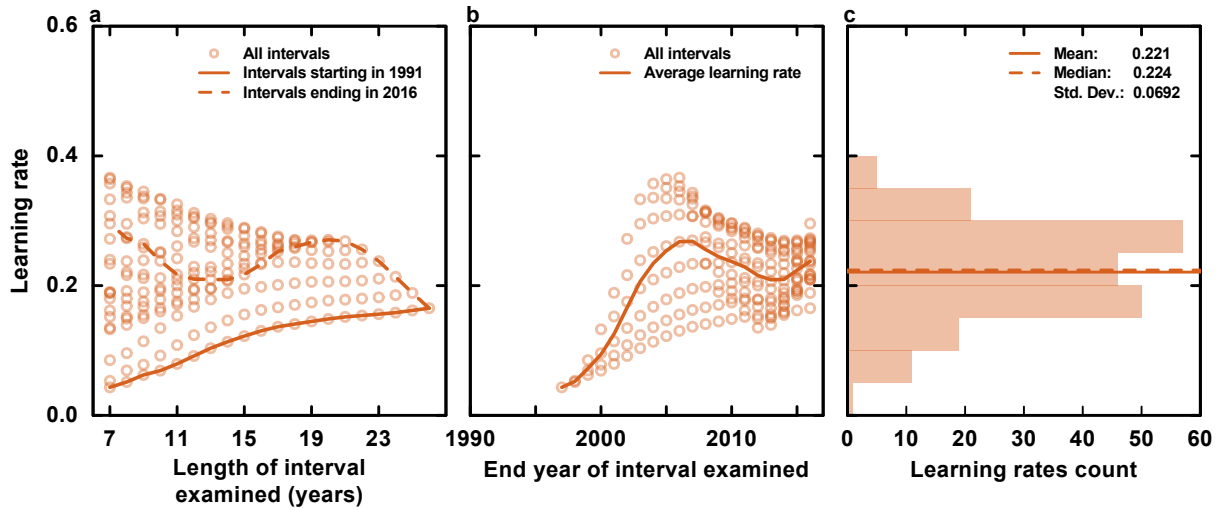

Figure 7: Learning rates calculable from the representative price and market size series. Learning rates calculated for every interval of seven or more years between 1991 and 2016 by regressing representative price per energy capacity against market size measured in energy capacity, plotted by interval length (a) or interval end year (b), along with a histogram of all improvement rates calculated (c). An alternative plot, where the dependent variable is the slope of the line fit to the logarithmized data, is also available (Figure S11).

For a given representative price series, learning rates obtained by regression versus cumulative market size measured in number of all types of cells (21.7 and 20.4% for cylindrical and all-cell-types prices, respectively) are slightly higher than those obtained from regression against market size measured in energy capacity (20.1 and 18.9%, for cylindrical and all-cell-types prices). This trend is observed across a range of possible learning rate estimates (Figure S12 versus Figure S13). These results suggest that incorporating change in cell energy capacity into the market size estimate leads to an underestimate in the learning rate. These results also indicate that regressing a price series specific for cylindrical cells against a market size series representing all types of cells yields a learning rate estimate nearly 4% lower than that calculated when regressing against a series reflecting only cylindrical cells. Both cases lead to underestimates of the rate of technological change upon growth in cumulative market size.

By developing representative series and applying various performance curve models, we find that a range of proposed determinants correlate reasonably well with the scaled real price decline of lithium-ion cells. Prices are generally more correlated with time and cumulative series, which inherently incorporate time, of market size and patent filings than annual value series. Moreover, while the scaled real prices of all types of cells and cylindrical cells declined at very similar rates over time, their rate of price decline versus

| Price series | Market size estimate | | |
|---|---|---|---|
| | Num. of cylindrical cells | Num. of all cells | Energy capacity (MWh) |
| Cylindrical | 0.240 (0.0108) | 0.217 (0.0092) | 0.201 (0.0091) |
| All types | NA | 0.204 (0.0140) | 0.189 (0.0131) |

Table 1: Learning rates and error ($\sigma_{IR}$) estimated using different combinations of representative price per energy capacity and cumulative market size series, for the period 1992 –2016

cumulative market size differed considerably. We also found that the time period examined can noticeably impact learning rate estimates, while using different market size units has a noticeable but smaller impact. The variability in these results can be used to inform appropriate ranges for projections of lithium-ion technology improvement and, along with model-specific errors estimates,[27] reduce the cost of uncertainty in energy technology forecasts.

## Incorporating other performance characteristics

So far, these analyses have explored improvement in the real price of lithium-ion technologies scaled by energy capacity, in units of 2018 USD per kWh. In this metric, the energy capacity represents the service provided by a lithium-ion cell. However, energy capacity is only one measure of a cell's performance. Other characteristics of lithium-ion cells, such as energy density (Wh/liter), specific energy (Wh/kg), power density (W/L), specific power (W/kg), cycle-life, self-discharge, temperature sensitivity, and safety, have long been the focus of considerable research and development efforts; and as a result many of these characteristics have improved substantially since the early 1990s.[17,32,81,94–103] Improvements in most of these characteristics were driven by cells' applications. For example, energy density is an important characteristic for small portable electronics while power density is more important for power tools and electric vehicles. Limiting the definition of service to only a cell's energy capacity ignores other changes in performance or quality, as has been observed for other technologies.[27,104–106]

We sought to explore the application of performance curve models of technological improvement and resultant changes in improvement rates when additional features of lithium-ion technology performance are considered. We are specifically interested in energy density and specific energy. Both characteristics have been and remain important features of lithium-ion technologies that enable their application to portable consumer electronics and transportation systems.[98,100,107] To expand the definition of unit service provided by a lithium-ion cell, we define its service as the product of its cell-level attributes ($g_i$) weighted by constants ($h_i$)

$$\text{Service per cell} = \prod_i g_i^{h_i} \tag{9}$$

where these attributes can be energy capacity per cell, energy density, cycle-life, etc. Using this formulation, the definition of service provided by a cell can be expanded to include energy density as:

$$\text{Service per cell} = \left(\frac{\text{energy capacity}}{\text{cell}}\right)^{h_1} \times (\text{energy density})^{h_2} \quad (10)$$

The resulting price per service equation is then formulated

$$\text{Price per service} = \frac{\text{Price / cell}}{(\text{energy capacity / cell})^{h_1}} \times \frac{1}{(\text{energy density})^{h_2}} \quad (11)$$

This definition of price per service limits price to currency-valued terms. Alternatively, one could define the price of a cell broadly, for example to include both a monetary price per energy capacity and volumetric price per energy capacity, as in:

$$\text{Price per service} = \left(\frac{\text{Price}}{\text{energy capacity}}\right)^{j_1} \times \left(\frac{\text{Volume}}{\text{energy capacity}}\right)^{j_2} \quad (12)$$

Volumetric price can be interpreted as the space in a mobile device or vehicle that must be available to accommodate the cell. A term combining these two prices could be constructed by their multiplication, yielding a result analogous to that expressed in eq. 11. While in either formulation specific energy could be similarly included as a third factor, energy density and specific energy are strongly correlated (see Figures S14 and S15) at the cell level. Thus, our modeling only considers one of these two performance metrics at a time.

The multiplicative form in equation 9 is similar to that employed in multiattribute utility theory,[108] among many other phenomenological 'two-factor' models. While this estimation of service does not rely on preferences obtained by interviewing cell manufacturers or purchasers,[109] its multiplicative form is sensible in this context. A cell with no energy capacity provides no service, regardless of its energy density. Similarly, a cell with high energy capacity but very low energy density, such as a large lead-acid cell, is similarly less useful for portable and transportation applications, which have driven the development and deployment of lithium-ion technologies over most of their history. Increasing either energy capacity or energy density for a given cost or price can be considered technological improvement. Without detailed survey data, assignment of values to the weighting constants ($h_n$, $j_n$) would be arbitrary, so for this study both are assumed to be equal to one, implying that energy capacity and energy density are considered equally important cell-level attributes and that these preferences have remained consistent over time. Setting the weighting constants to one in either eq. 11 or eq. 12 also provides a physically reasonable relationship between service and energy capacity and allows price per service to be estimated using contemporaneous cell-level price per energy capacity time series and energy density time series. This functional form is proposed as a first-pass model for adjusting cost for other aspects of technology performance.

To determine how energy density and specific energy of lithium-ion technologies improved over time, we collected records of lithium-ion cells between 1990 and 2019. Over this period, commercially available cells' maximum energy density (Figure 8) and specific energy (Figure S16) increased considerably. Diversification of these characteristics was also observed; many cells had energy densities and specific energies lower than the highest achievable at a given time. A variety of approaches were considered to develop series to represent how these characteristics changed over time (Figures S17 and S18), and series that tracked the 98$^{th}$ percentile annually were chosen (Figures 8 and S16). Series that tracked annual maxima or prevented decreases were rejected as they gave too much weight to individual data points or years, respectively. Average energy density and specific energy series were also considered because the representative price series comprise average prices. However, the data necessary to weight performance characteristics by market share were not available.

These data show that from 1991 through 2018, achievable energy density rose from approximately 200 Wh/l to over 700 Wh/l while specific energy rose from approximately 80 Wh/kg to over 250 Wh/kg. For both metrics, the series representing all cell types is similar to that for cylindrical cells as our data indicate that cylindrical cells tend to have the highest annual energy density and specific energy values. These series estimate how technological capabilities changed over time and do not reflect the how the market shares of cells with different performance characteristics might have changed.

Given the representative price per energy capacity and energy density series, an annual price per service series can be calculated for both all types of cells and cylindrical cells, where service is defined in equation 10 and simplified to give a price per service series as defined in equation 11. Then, the aforementioned performance curve models (eq. 1 through 5) can be applied to relate this price per service series to possible determinants and examine how the empirical relationships change when the definition of service is expanded to include both energy capacity and energy density. As was observed when examining price per energy capacity, application of equations 1, 2, and 4, reveals reasonably high correlations between the all-cell-shapes representative price series and time, cumulative market size, and cumulative patent filings (Figure 9). However, in all cases the slopes of the linear models are considerably steeper when service includes energy density in addition to energy capacity, suggesting that lithium-ion technologies improved more rapidly than estimated from price per energy capacity measures alone. In the case of equation 1, considering energy density as part of service results in an annual percent decline in price per service of 17.1% for all cell types, markedly higher than that observed for price per energy capacity (13.1%). The learning rate for all cell types similarly increases from 20.4 to nearly 26.6% while the inventive activity rate increases from 40.1% to 49.7%. Including energy density within the scope of service also increases improvement rates calculated when applying equations 3 (Figure S19) and 5 (Figure S20) to the series representing all types of cells. When service includes energy capacity and specific energy, as opposed to energy density, similar increases in rates are observed (Figures S22–S24).

Given price per energy capacity and energy density series specific to cylindrical cells, equations 1–3 can also be used to examine how incorporating energy density into the definition of service impacts rates determined for the cylindrical cells subgroup (Figure S21). With all three models, similarly high correlations are observed regardless of how service is defined while the slopes of the trend lines are considerably steeper when energy density is incorporated. In the case of equation 1, the annual decrease in price per service over time increases from 13.3 to 17.4% upon incorporation of energy density, nearly the same as the increase observed for all cell types. Meanwhile, the learning rate increases from 24.0 to 30.9%, which is slightly larger than the increase observed for all cell types. Slightly smaller increases in rates were obtained when service is defined as the product of energy capacity and specific energy (Figures S25).

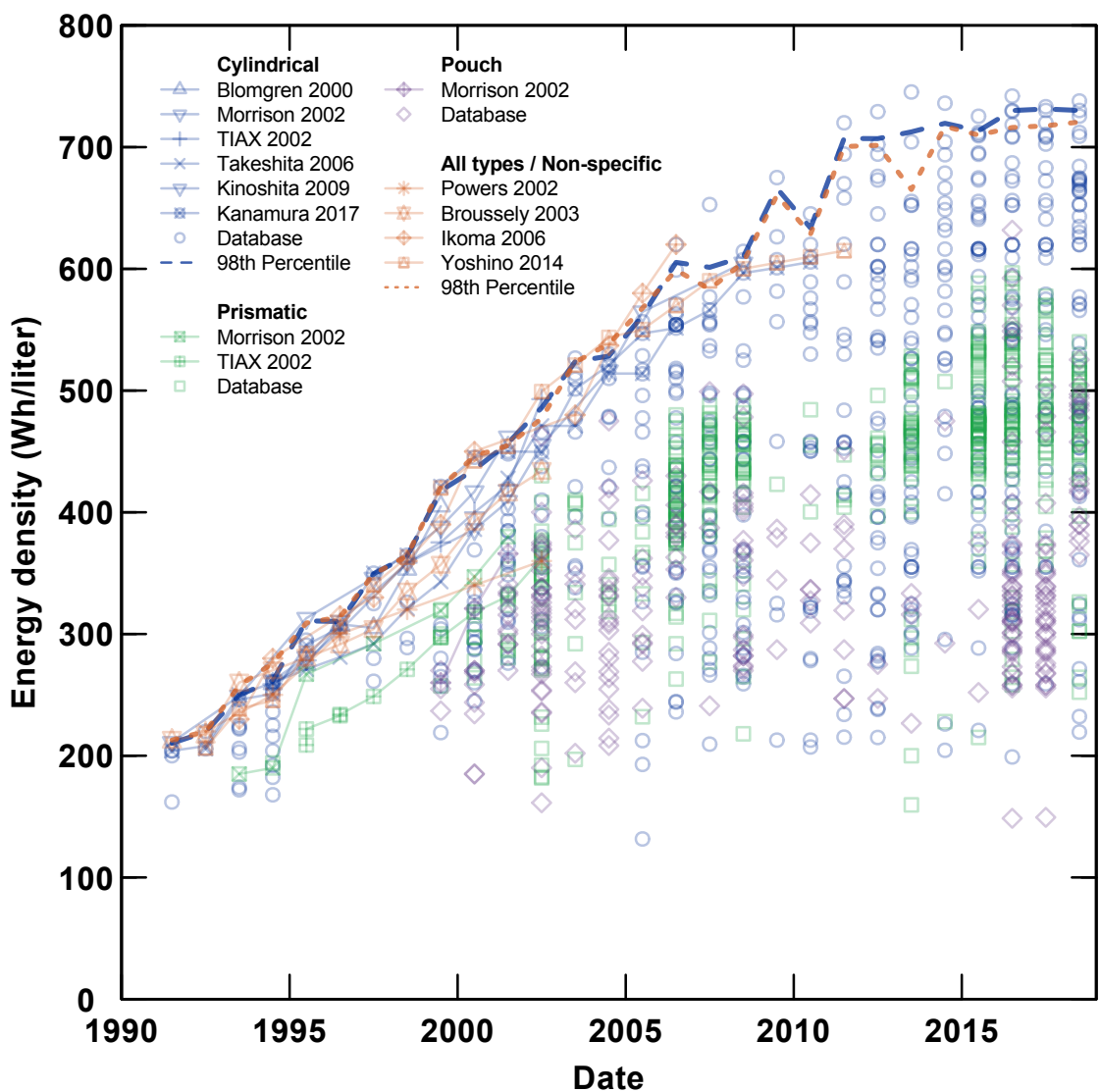

Figure 8: Lithium-ion cell energy density over time. Time series and single-year records of nameplate energy density values for lithium-ion cells for cylindrical (blue), prismatic (green), pouch (purple), and all types (orange) of cells, as well as representative price series for cylindrical (blue, bold, dashed) and all types (orange, bold, dashed) of cells. Series that did not specify cell type are included with series representing all types of cells. An analogous plot for specific energy values is included as Figure S16.

In addition, the relative correlations observed for all types of cells versus the cylindrical subgroup are maintained when the definition of service is expanded to include energy density (Figure 10). Specifically, the correlation of price per service with time is slightly higher for the all-cell-types series, while the correlation with cumulative market size is modestly higher for the cylindrical cells subgroup.

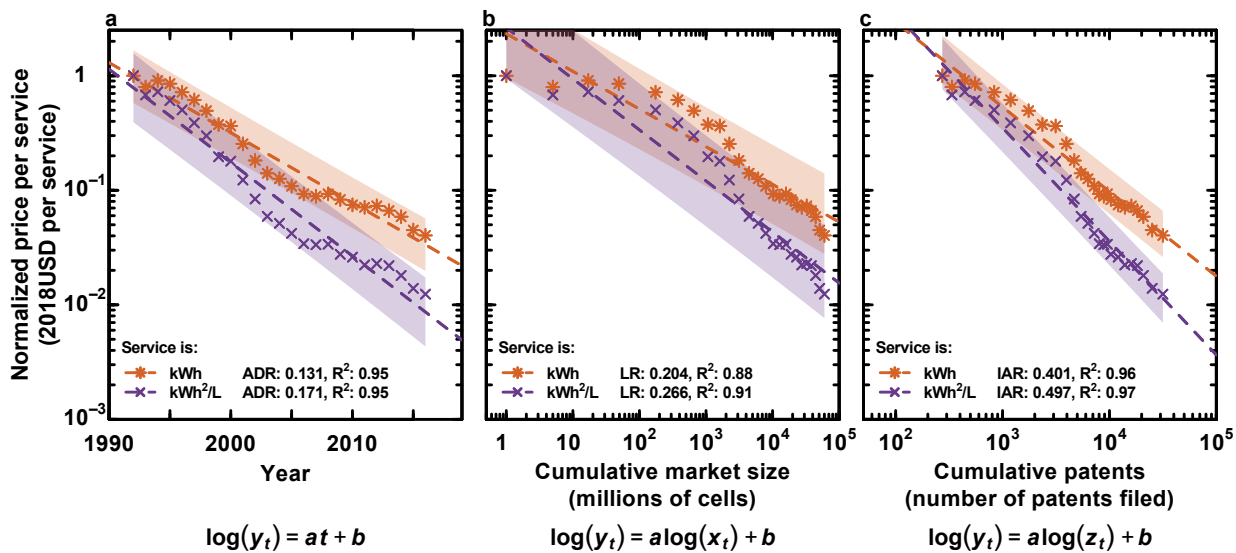

Figure 9: Lithium-ion price per energy capacity and per service regressed against a variety of determinants for all cell types. Lithium-ion cell price per energy capacity (orange asterisk marks) and price per service (purple x marks) regressed against year (a), cumulative market size (b), and cumulative patent filings (c) for all cell shapes. Prediction intervals (95% level) are plotted as similarly colored shaded regions. Analyses are restricted to the years 1992 through 2016, and market size is measured in number of cells.

## Discussion

This analysis combines data from and reconciles differences between 90 series that describe how lithium-ion technologies have changed and possible drivers of that change. Representative series that track changes in price, market size, patent filings, and cell-level energy density and specific energy were constructed for all types of lithium-ion cells and in most cases also for cylindrical cells, allowing us to compare trends in this important subgroup to those observed for all cell shapes. By combining and harmonizing data from a variety of sources, we sought to develop more reliable estimates of technological change and improvement rates for lithium-ion technologies. Moreover, by clearly delineating how these representative series were constructed, we aim to provide a methodological framework that can be extended, both as additional data on lithium-ion technologies are collected and to other technologies.

We used performance curve models to examine how the real price of lithium-ion cells changes with time,

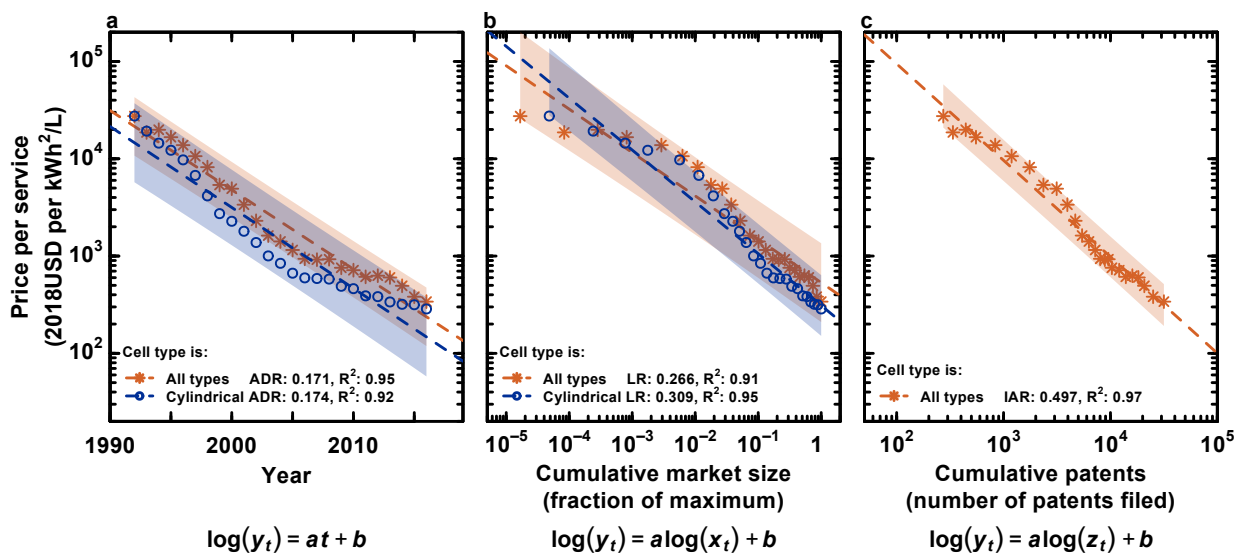

Figure 10: Lithium-ion price per service regressed against a variety of determinants. Lithium-ion cell price per service (in kWh$^2$/L) regressed against year (a), cumulative market size (b), and cumulative patent filings (c) for all (orange asterisk marks) and cylindrical (blue circles) cell shapes. Prediction intervals (95% level) are plotted as similarly colored shaded regions. Analyses are restricted to the years 1992 through 2016, and market size is measured in number of cells produced.

cumulative market size, and cumulative inventive activity for the period from 1992 through 2016. In the first case, modeling energy capacity–scaled real prices as decreasing exponentially with time, we observed similar annual price decreases for both all types of cells (13.1%) and cylindrical cells (13.3%) (Figure 11a). These rates are just below the mean (13.7%) and median (13.6%) of the annual decrease percentages calculated for the price series collected in this work (cf. Figure 1). In addition, these rates suggest prices declined more rapidly than was observed by Anderson for lithium-ion technologies (9.9% for 1998–2005, 5.4% for 2002–2005)[61] and are similar to the rate reported by Deutsche Bank analysts (14% for "laptop battery costs").[110,111] The rates are also more rapid than the rate Koh and Magee estimated for a range of energy storage technologies (ADR: 3.1% for USD per Wh, as transformed from their "annual progress" exponential coefficient for stored energy per unit cost).[32] The low rate observed by Koh and Magee likely results from their cost change analysis relying primarily on lead-acid technologies. In addition, the annual decrease percentages we estimate for lithium-ion technologies are faster than the average annual decrease percentage measured for many other industries (7.6%) (Figure 11a).[27] Specifically lithium-ion technologies have undergone a greater annual percent decline than the average observed for a range of chemical technologies (6.1%) and energy technologies (4.8%).

Price per energy capacity also declines with cumulative market size as measured in number of cells,

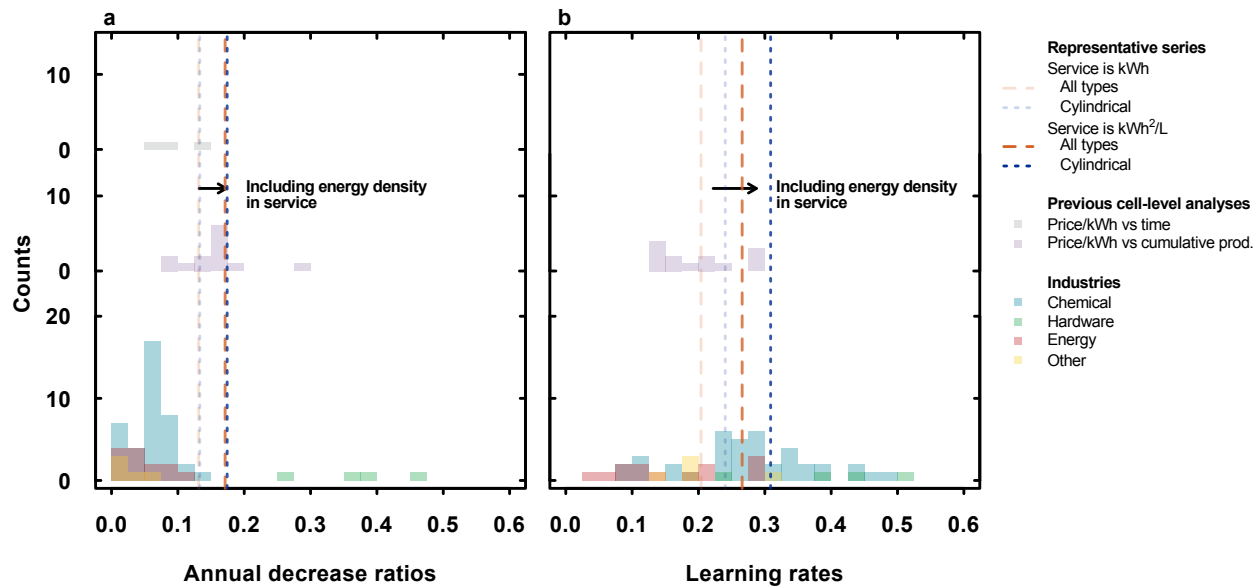

Figure 11: Annual decrease ratios (ADRs, (a)) and learning rates (LRs, (b)) for lithium-ion cells and other industries. Ratios and rates estimated in this work are denoted with vertical dashed lines for all types of cells (orange) and cylindrical cells (blue), with service defined as real price scaled by kWh (faded) and $kWh^2/L$ (bolded). Previous analyses' estimates of lithium-ion cells' price decline versus time (gray) and cumulative production (purple) are plotted in the top two histograms. Analyses versus time provide comparable ADRs while those versus cumulative production provide both ADRs and LRs. (Details on specific analyses are available in Table S2.) The lowermost histograms summarize the ADRs and LRs estimated for a range of technologies, grouped by industry type, as reported previously.[27] The ostensible outlier among the previously reported lithium-ion ADRs relies on a short data series representing only cells destined for automotive applications.

with estimated learning rates of 20.4% for all types of cells and 24.0% for cylindrical cells (Figure 11b). These rates are faster than those calculated when regressing price per energy capacity against cumulative production in MWh (18.9%) as determined herein, as well as those rates Nagelhout et al. estimated (17%)[45] and Schmidt et al. found specifically for 18650-sized cells (19 ± 3%).[14] The learning rate for all types of cells is also just below the mean (20.7%) and above the median (19.0%) of the previously reported learning rates for lithium-ion cells, while the rate for cylindrical cells is considerably above both. All of the learning rates determined in this analysis are between the rate recently estimated by Kittner and coworkers (15%)[12] and the central cell-level rate employed by Schmidt and coworkers (30%).[14,46] In addition, these learning rates are well within the ranges observed previously for a variety of technologies,[27,93] and specifically are above average compared to those observed for energy technologies (17%) but below the average estimated for a variety of chemical production technologies (28%) (Figure 11b).

Modeling annual energy capacity–scaled prices as a function of cumulative patent filings exhibits the highest coefficient of determination ($R^2$), as was observed by Kittner and coworkers.[12] However, our results reveal a steeper slope, estimating a doubling of cumulative patent filings is associated with a reduction in price of 40.1%, compared to Kittner and coworkers' estimate of 31%.[12]

Higher correlations, as indicated by their coefficients of determination, are observed between lithium-ion technologies' price per service and cumulative measures of market size and inventive activity than between price and their annual measures. This difference could result from cumulative measures inherently incorporating time-dependent factors, such as research and development and learning-by-doing, along with the economies of scale that are reflected in annual market growth.[27,90]

A variety of factors have contributed to the diversity of previously reported annual decrease ratios and learning rates. Notably, many of the learning rates estimated by regressing price per energy capacity versus cumulative production fall within the envelope of rates calculable from consecutive subsets of representative price per energy capacity and market size series, suggesting that some of the variability in reported rates could result from the various time periods considered by different researchers. Additional variability results from data treatment choices, such as whether previous analyses included inflation correction and whether a price series specific to cylindrical cells was regressed against market size estimates for all types of cells. Meanwhile, changing the market size measure from energy capacity to numbers of cells only slightly increases improvement rate estimates.

Grouping data by cell type reveals notable differences between the modeling and measured improvement rates for all types of cells and cylindrical cells. When examining decline in price, regardless of how service is defined, we observe a higher correlation between cylindrical cells' price decline and cumulative market size than between all cell types' price decline and cumulative market size. However, cylindrical cells' price decline is less correlated with time than the all cells types' price decline is. In addition, the learning rate for cylindrical cells is greater than that observed for all types of cells, while both groups have very similar annual decrease ratios. These differences in correlation and improvement rates between the whole group (all cell types) and subgroup (cylindrical cells) suggest that switching between cell types has allowed lithium-ion technologies to improve generally with time while benefits from learning-by-doing and economies of scale are more pronounced when analysis is limited to a single technological design. As more data become available these hypotheses can be further investigated.

We also introduce a method to expand the definition of service provided by lithium-ion cells to improve estimates of how their overall performance has improved. While a price per energy capacity metric for energy storage technologies presents a convenient analogue to the cost per installed power capacity metric commonly used to estimate learning rates for electricity generating technologies,[33,35] its use belies the fact that while electricity supplied to a grid is generally fungible, especially when supplied reliably or on-demand, electrochemical cells are not. For example, a less expensive but considerably larger and heavier battery technology would not likely replace lithium-ion cells in most portable electronics. The need to and difficulty of incorporating additional performance metrics has long been a challenge for phenomenological studies of technological progress.[53] Even Wright noted in his seminal study of airplane costs that "time saving" was a difficult-to-value metric required to compare travel in a plane to that in a car.[29] Moreover, researchers have found that when preferences for a given technology's performance characteristics change, deviations from the classic power law relationship between price per unit and cumulative production can be observed, as was the case when Ford shifted focus from producing increasingly inexpensive automobiles to improving characteristics including comfort, performance, and safety.[52,112] In the case of Ford's transition from the Model T to the Model A, measures of price per vehicle and price per pound did not reflect the value of improvements in these other characteristics in their definitions of service (*i.e.* "vehicle" or "pound").

When we include energy density or specific energy in the definition of service to better estimate the overall improvement rates for lithium-ion technologies, we measure much faster rates of technological change than were observed for price per energy capacity alone. For the period from 1992 through 2016, when service includes energy capacity and energy density, annual percent decreases in price per service increase considerably, from 13.1 and 13.3% to 17.1 and 17.4% for all types of cells and cylindrical cells, respectively (Figure 11a). Similarly, learning rates also increase as the definition of service expands, to 26.6 and 30.9% for all types of cells and cylindrical cells, respectively (Figure 11b). When regressing against cumulative patent filings, the inventive activity rate increases from 40.1% to 49.7%. Greater rate increases are observed when energy density is incorporated than when specific energy is, reflecting the larger relative gains observed for energy density improvements since the early 1990s.

The increase in improvement rates observed upon incorporating other important metrics suggests the degree to which rates measured from only price per energy capacity series underestimate how rapidly lithium-ion technologies have improved. They similarly provide a rough approximation of how much a focus on non-cost performance characteristics might have limited cost or price decline. As different performance characteristics might be prioritized in the future, incorporating additional relevant characteristics into the definition of price per service could enable more accurate measures, and possibly projections, of technological change, though further research is needed. For example, the requirements of stationary storage applications have already started shifting focus from energy density and specific energy metrics to a variety of other characteristics, such battery lifetime and degradation.[2,9,20,22,46,107,113] Such cycle-life characteristics were actually incorporated into definitions of service early in the development of lithium-ion technologies.[99] Notably, a few researchers included cycle-life in their comparisons of lithium-ion cells to other battery technologies, summarizing service as "accumulated discharge energy" or the product of energy capacity and cycle life, sometimes corrected for capacity loss over time.[20,81,97,100,114] Metrics including cycle-life could provide a better estimate of how lithium-ion technologies have improved or could improve with respect to stationary storage applications. While a dearth of reliable, comparable historical records on capacity fade in lithium-ion cells complicates retrospective analysis, prospective use of this type of metric could aid technology comparisons and projections of cost and price decline. However, additional research is needed on cycling and capacity-loss characteristics across applications.

Advancing the study of technological innovation requires careful data collection,[59,115] consideration of data uncertainties, and examination of different technologies.[27] Strengths of this work include systematic, careful collection, harmonization, and combination of data that describe how lithium-ion technologies evolved, and we methodically detail our approach to data collection and analysis and the methods we employ to provide a blueprint for others who seek to perform similar analyses. Notably, we sought to trace data as far as possible to their original sources to explicate various conversions and assumptions, and if available, tried to compare data from different original sources. Moreover, we carefully differentiate between data that referred to either all types of cells or only cylindrical cells when constructing time series and estimating improvement rates. In addition, we expand the definition of service to include other important technology characteristics in order to better estimate how rapidly lithium-ion technologies have changed, and to begin to understand how they might change in the future.

A key limitation of this study is its incorporation of data with unknown original sources and collected with sometimes unclear methods. Notably, some of the price and market size series rely in part on data reported by industry consultants, whose data collection could involve a variety of methods or assumptions that are not always presented with the final data. Comparison with government-provided data, especially those provided by Japan's METI, and a variety of other sources helped mitigate this weakness. However, the possibility remains that the consultants themselves used each other's data or the government data to develop their data series, which could create the appearance of more independent data series than actually exist. We worked to address these issues by transparently presenting the methods employed to construct the aforementioned representative series so that these approaches and analyses can be improved as more data become available. The first step toward addressing these issues is to elucidate them. We also expect these methods could aid those dealing with similar sources of data uncertainty inevitably encountered when studying technological change and reduce the cost of this uncertainty through characterizing it such that it can be incorporated into an error model to be used for projections.[27] Another limitation is that our analysis focused on all cell types and cylindrical cells, as insufficient data were available to confidently provide similar results for prismatic and pouch cells. Data availability further constrained many of our analyses to the period from 1992 through 2016. Finally, our expansion of the definition of service was limited to incorporating energy density and specific energy performance metrics.

## Concluding remarks

Based on a thorough examination of available data, this work provides rigorous estimates of the decline in price, growth in market size and inventive activity, and improved technical performance of lithium-ion technologies. Using performance curve models, we report robust data detailing and estimates of the rate of lithium-ion technologies' advancement. We found that while the prices of both cylindrical cells and all types of cells declined similarly with time, a considerably higher learning rate is observed for cylindrical cells. To expand our measures of lithium-ion technologies' change, we propose a model to incorporate other attributes into the definition of service provided by a lithium-ion cell and find that when energy density or specific energy are included improvement rate estimates increase considerably.

The increase in improvement rates observed when other historically important performance characteristics are incorporated into the definition of service suggests a rough estimate for how much measures based on price per energy capacity alone might underestimate the how rapidly lithium-ion technologies improved. The increase similarly gives an approximate indication of how much price decline might have be limited by a focus on these performance characteristics. As the requirements for these performance characteristics are relaxed, as in the case of stationary storage applications, priorities for research, development, and production efforts are expected to transition. As a result, cost or price for a different service may decline more rapidly than would be suggested by rates that only consider price per energy capacity. However, engineering-based mechanistic modeling of lithium-ion technologies' historic and possible future cost change is required to further evaluate this potential.

Measuring technological change often requires working with limited data sets that contain measurement and sampling uncertainty. Our data collection and analysis approaches aim to further delineate a model for performance curve analyses and highlight methods for additional discussion and improvement. We expect that the methodology presented herein and the approach of incorporating important intensive characteristics into broader cost or price per service metrics could be applied to a range of technologies and help improve measures and projections of technological change.

## Conflicts of interest

The authors declare no competing financial interests.

## Acknowledgments

We thank the Alfred P. Sloan Foundation for funding this research. We also thank Dr. James McNerney, Dr. Goksin Kavlak, and Dr. Barry I. Graubard for useful conversations and advice as well as David Morrison for copies of conference notes.

## Notes

[a]We refer to these relationships and others that relate a performance measure to an experience measure generally as "performance curves".[52,116]